\documentclass[aip,apl,twocolumn,amsmath,amssymb,bibnotes]{revtex4}

\usepackage{color,epsfig,rotating,graphicx,subfigure,}

\usepackage{amsmath,amssymb,amscd,ulem}

\usepackage[latin1]{inputenc}
\usepackage[english]{babel}

\usepackage{dsfont}
\renewcommand{\Re}{{\rm Re}}

\newcommand{\re}{{\rm e}}
\newcommand{\rd}{{\rm d}}

\newcommand{\rs}{{\rm s}}
\newcommand{\rp}{{\rm p}} 
 
\newcommand{\kb}{k_{\rm B}}

\begin{document}
\title{Harvesting the electromagnetic energy confined close to a hot body}

\date{\today}

\author{Philippe Ben-Abdallah$^1$ and Svend-Age Biehs$^{2}$}
\affiliation{$^1$ Laboratoire Charles Fabry,UMR 8501, Institut d'Optique, CNRS, Universit\'{e} Paris-Sud 11,
2, Avenue Augustin Fresnel, 91127 Palaiseau Cedex, France}
\affiliation{$^2$ Institut f\"{u}r Physik, Carl von Ossietzky Universit\"{a}t, D-26111 Oldenburg, Germany}

\begin{abstract}  
In the close vicinity of a hot body, at distances smaller than the thermal wavelength, a high electromagnetic energy density exists due to the presence of evanescent fields radiated by the partial charges in motion around its surface. This 
energy density can surpass the energy density in vacuum  by several orders of magnitude. By approaching a PV cell with a 
band gap in the infrared frequency range this non-radiative energy can be transferred to it by photon tunneling and 
surface mode coupling. Here we review the basic ideas and recent progress in near-field energy harvesting.
\end{abstract}


\maketitle

\section{Introduction}

Thermophotovoltaic devices~\cite{Coutts,Lenert} are energy conversion systems that generate electric power directly from thermal radiation. 
While in classical photovoltaic conversion devices the efficiency (defined as the ratio of the electric power produced 
by the cell over the net radiative power exchanged between the primary source and the cell) is bounded by the thermodynamic 
Shockley-Queisser limit~\cite{Shockley} (30\% for cells with a gap energy $E_g=1\,{\rm eV}$ and 10\% for $E_g=0.5\,{\rm eV}$), with near-field 
thermophotovoltaic (NTPV) systems this limit can, in principle, be surpassed by exploiting the strong heat transfer 
due to tunneling of non-propagating photons. In particular, with a quasi-monochromatic source, as those made with materials 
which support a surface wave, the efficiency can be close to one when this frequency coincides with the gap frequency of 
semiconductor~\cite{Laroche}. Moreover, the strong magnitude of exchanges in plane-plane geometry~\cite{PvH} in near-field regime compared 
with what happens in far-field regime (which cannot exceed the blackbody limit~\cite{Planck}) can lead to a generated power 
between 1 W\,cm$^{-2}$ and 120 W\,cm$^{-2}$ in the temperature range 500K-1200K for separation distances between the source and 
the cell of 10nm~\cite{Ilic}. As a result a cell of 25cm$^2$ could theoretically generate a power of 25W-3000W. This is a very large value.
The typical energy demand of a household in US~\cite{US20} is 2500W, for instance.
 
In this paper we describe the basic principles behind this near-field technology and we briefly discuss the main issues which limit 
to date its massive deployment.

\section{Electromagnetic energy close to a hot body}

It is well known that the spectral energy density of thermal radiation at a given equilibrium temperature $T$ follows the Planck law
\begin{equation}
  u_{{\rm BB},\omega} = \Theta(T) D_{\rm vac}(\omega)
\end{equation}
where 
\begin{equation}
  \Theta(T) = \frac{\hbar \omega}{\re^{\hbar \omega / \kb T} - 1}
\end{equation}
is the mean energy of a harmonic oscillator in thermal equilibrium and 
\begin{equation}
  D_{\rm vac} (\omega) = \frac{\omega^2}{\pi^2 c^3}
\end{equation}
is the density of states in vacuum. Here, we have introduced the Boltzmann constant $\kb$, the reduced Planck constant $\hbar$, 
and the light velocity in vacuum $c$. The maximum of the Planck spectrum is at a frequency of $\omega_{\rm th} = 2.82 \omega_c$
with $\omega_c = k_{\rm B}T/\hbar$. As function of wavelength the maximum is at the thermal wavelength $\lambda_{\rm th}$
given by Wien's displacement law $\lambda_{\rm th} = 2897\mu{\rm m}\,{\rm K}/T$, i.e.\ at $300\,{\rm K}$ the
dominant frequency is $\omega_{\rm th} \approx 10^{14}\,{\rm rad/s}$ and the dominat wavelength $\lambda_{\rm th} \approx 10\,\mu{\rm m}$.

\begin{figure}
  \includegraphics[ width = 0.3\textwidth]{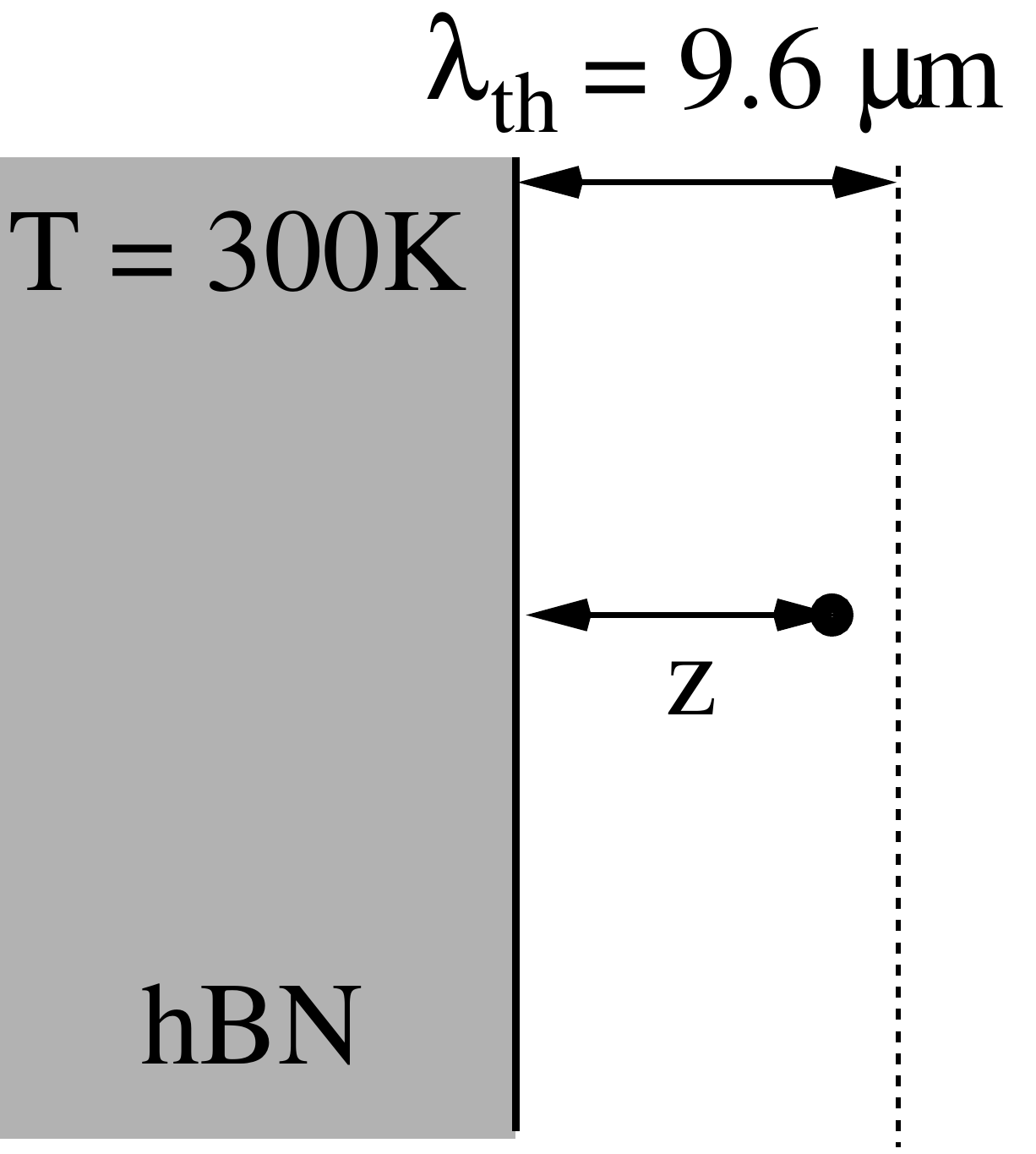}
  \caption{Sketch of the planar sample in global equilibrium at temperature $T = 300\,{\rm K}$. The energy density at a near-field distance $z < \lambda_{\rm th} = 9.6\,\mu {\rm m}$ highly depends on the properties of the thermal emitter which is here hBN.\label{Fig:UBB}}
\end{figure}

Now, close to a thermal emitter at distances smaller than the thermal wavelength $\lambda_{\rm th}$ the properties of the
emitter itself enter into the density of states and therefore also into Planck's law. Close to a thermal emitter as sketched in Fig.~\ref{Fig:UBB} in the particular case of a hexagonal boron nitride (hBN) sample a 
generalized Planck law can be derived which has the form~\cite{Agarwal,Eckhardt,Dorofeyev}  
\begin{equation}
  u_\omega = \Theta(T) D_{\rm loc}(\omega)
\end{equation}
where $D_{\rm loc}(\omega)$ is the local density of states. It depends in general on the geometry and optical properties of the
emitter and on the position with respect to the emitter at which it is evaluated. For an emitter with a planar interface due to
the translation symmetry with respect to planes parallel to the interface it depends on the distance $z$ and the properties of 
the emitter. It can be shown that for $z \gg \lambda_{\rm th}$ it converges to the density of states in vacuum $D_{\rm loc} \rightarrow D_{\rm vac}$
and in the opposite quasi-static regime for $z \ll \lambda_{\rm th}$ it becomes larger than the vacuum density of states and 
diverges like $D_{\rm loc} \propto 1/z^3$. Therefore, the energy density of thermal radiation close to a hot body is in general
distance dependent and it's magnitude can be larger than that predicted by Planck's law. This can be attributed to the contribution of evanescent waves existing in the vicinity of the emitter which have electromagnetic fields which are exponentially 
decaying with respect to the emitters interface. Such waves are 
for example total internal reflection waves and surface waves. As was already shown by Eckhardt~\cite{Eckhardt} and later studied in 
detail by Shchegrov {\itshape et al.} and Joulain {\itshape et al.}~\cite{Shchegrov,Joulain} the surface waves can lead to an 
extremely enhanced quasi-monochromatic near-field energy density spectrum. In metals such surface waves are surface-plasmon 
polaritons and in polar dielectrics such surface waves are known as surface-phonon polaritions (SPhP). 

\begin{figure}
  \includegraphics[ width = 0.45\textwidth]{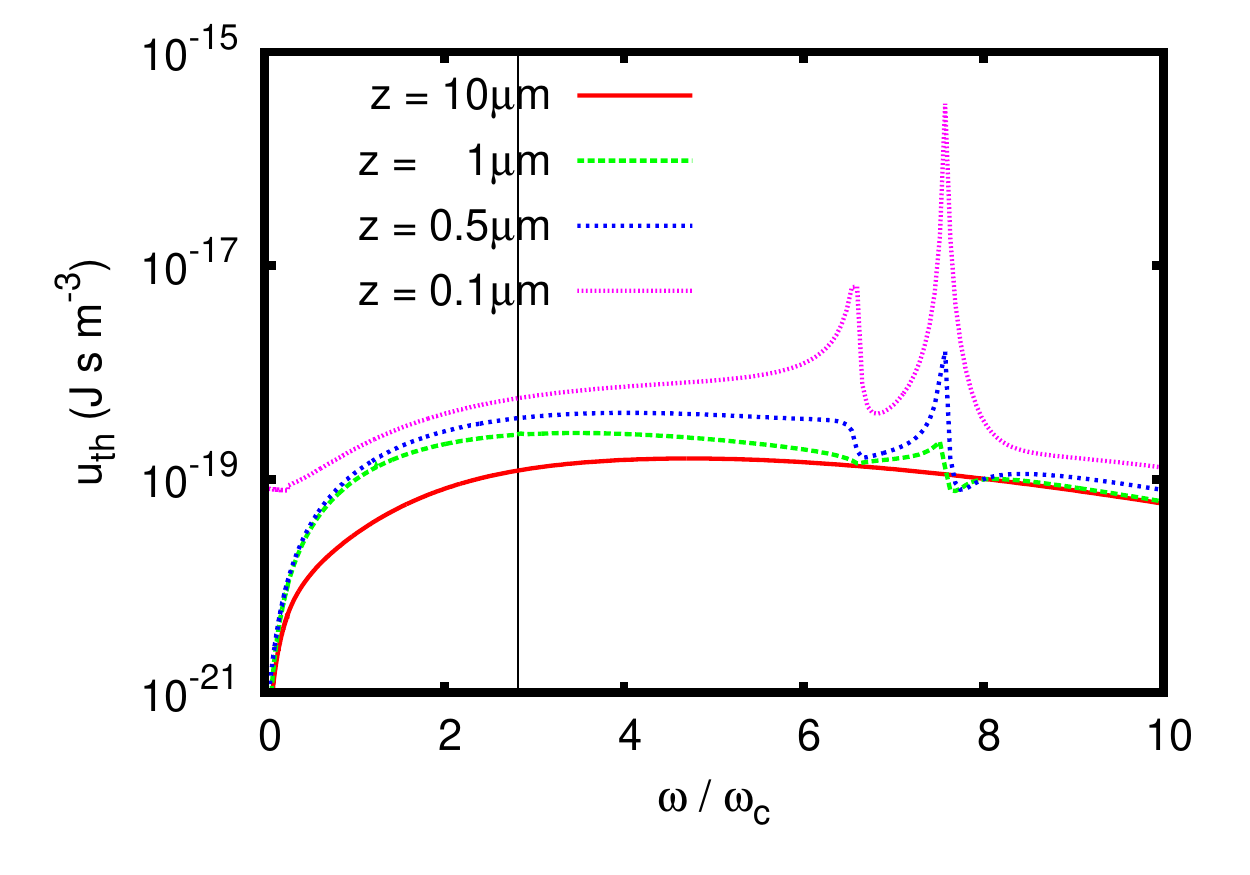}
  \caption{Semi-log plot of the spectral energy density $u_\omega$ in a distance $z$ above a hBN halfspace at $T = 300\,{\rm K}$. The thermal wavelength is therefore $10\mu{\rm m}$. Therefore the spectrum at $z = 10\mu{\rm m}$ concides mainly with the blackbody spectrum $u_{{\rm BB},\omega}$. The vertical line is at the maximum frequency of the blackbody spectrum at $2.82 \omega_{\rm c}$ with $\omega_{\rm c} = 3.5\times10^{13}\,{\rm rad/s}$. For more details see Refs.~\cite{Dorofeyev,Shchegrov,Joulain}.~\label{Fig:Energydens}}
\end{figure}

In Fig.~\ref{Fig:Energydens} we show the energy density close to a hBN emitter. The optical phononic properties of hBN is described by the Drude-Lorentz model~\cite{Palik}
\begin{equation}
  \epsilon_{\rm e} (\omega) =\epsilon_{\infty}\frac{\omega ^2-\omega_\text{L}^2+i\gamma\omega}{\omega^2-\omega_\text{R}^2+i\gamma\omega},
\label{Fig:hBN}
\end{equation}
with $\epsilon_{\infty}=4.88$, $\omega_{\rm L}=3.032\times 10^{14}\,$rad/s, $\omega_{\rm R}=2.575\times 10^{14}\,$rad/s and $\gamma=1.001\times 10^{12}\,$rad/s. 
As shown in Fig.~\ref{Fig:Energydens} for hBN which supports surface-phonon polaritons the spectral energy density becomes dominated
by the surface-phonon-polariton resonance at the frequency $\omega_{sp}$ for very small distances $z \ll \lambda_{\rm th}$. 
This resonance frequency is in general for surface modes on planar interfaces implicitely defined by the condition $\Re(\epsilon(\omega_{sp})) = -1$ where $\epsilon$ is the complex permittivity of the emitter which here has an interface with the vacuum. For metals the surface mode resonances are typically 
in the UV region and for polar dielectrics they are in the infrared. In particular, for hBN the resonance frequency is $\omega_{sp} = 2.960\times10^{14}\,{\rm rad/s}$ which corresponds to a wavelength of $5 \mu {\rm m}$. Hence the resonance is very close to the thermal wavelength $\lambda_{\rm th}$ at $300\,{\rm K}$ and can therefore be thermally excited at room temperature which is not the case for the surface waves in metals. Note, that the thermal near-field energy density can be tuned by texturing the material or by deposing 2D sheets on its surface~\cite{pba_apl,Messina_prb} and  it is accessible with different scanning probe devices~\cite{DeWilde,DeWilde2,Kittel,Hillenbrand,Raschke,Komiyama,Komiyama2}.

\section{Near-field heat transfer}

\begin{figure}
  \includegraphics[ width = 0.4\textwidth]{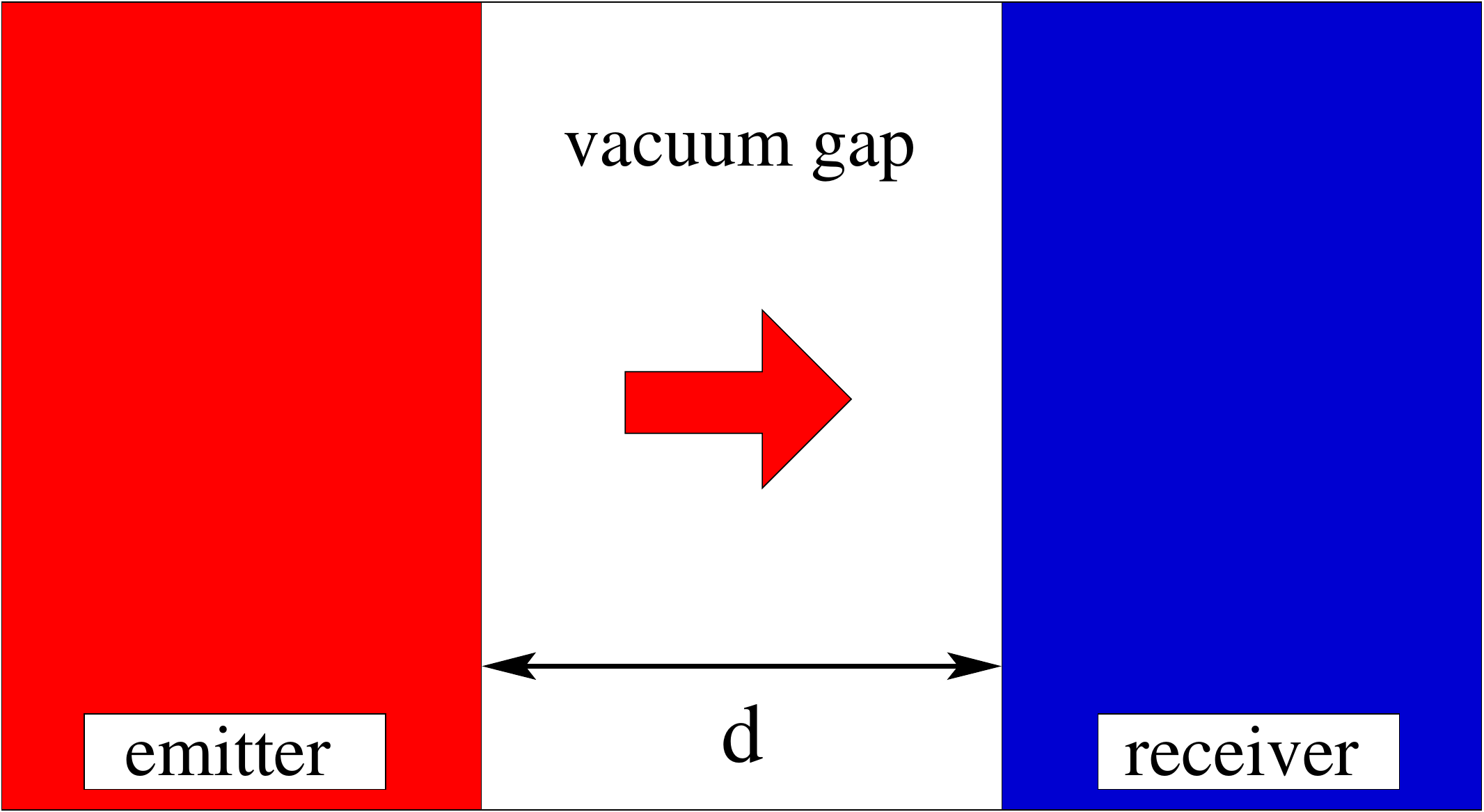}
  \caption{Sketch of the emitter and receiver separated by a vacuum gap of thickness $d$. In NTPV the receiver is replaced by a low band-gap photovoltaic cell. \label{Fig:Flux}} 
\end{figure}

When bringing a receiver with temperature $T_{\rm r}$ close to an emitter at higher temperature $T_{\rm e} > T_{\rm r}$ as sketched in Fig.~\ref{Fig:Flux} then in general the 
propagating and evanescent waves generated by the thermal motion of the charges inside the emitter and receiver will contribute 
to the heat transfer. It has already been shown in the 1970's that when the receiver enters the near-field regime of the emitter, i.e.\
for emitter-receiver distances $d < \lambda_{\rm th}$ then the resulting radiative heat flux $\Phi$ can in this case 
become much larger than the blackbody value due to the contribution of the evanescent waves which are confined at the surface of 
the emitter when the distance becomes smaller than the dominant thermal wavelength~\cite{Carvalho,Tien,Domoto,PvH}. About thirty years later, the idea to exploit the huge energy density of the evanescent waves for energy harvesting was brought forward by replacing the receiver by a photo-voltaic cell~\cite{DiMatteo,Whale,Pan,Whale2}. In Ref.~\cite{Whale} a first detailed study of NTPV cells based on the framework of fluctuational electrodynamics~\cite{Rytov} is provided. A first experimental 
proof of concept was provided by DiMatteo {\itshape et al.}~\cite{DiMatteo2}. These works paved the way for a large number of mainly theoretical 
works on near-field thermo-photovoltaics. 

The theoretical foundation is the general expression for the heat flux between two materials with planar interfaces which was first 
determined within the framework of fluctuational electrodynamics~\cite{Rytov} by Polder and van Hove~\cite{PvH}. 
This expression can be cast into a Landauer-like form by~\cite{LandauerPBA,LandauerSAB}
\begin{equation}
    \Phi(d)  = \int_0^\infty \!\! \frac{\rd \omega}{2 \pi} \Delta \Theta(\omega) \sum_{j = \rs,\rp}\int \!\! \frac{\rd^2 \mathbf{k}_\perp}{(2 \pi)^2} \, \mathcal{T}_j (\omega, \mathbf{k}_\perp;d)  
\end{equation}
where
\begin{equation}
   \Delta \Theta (\omega) = \Theta(T_{\rm e}) -  \Theta(T_{\rm r}) 
\end{equation}
and $\mathcal{T}_j (\omega, \mathbf{k}_\perp;d) \in [0:1]$ is the transmission factor for a given transversal mode with lateral wave 
vector $\mathbf{k}_\perp = (k_x, k_y)^t$, frequency $\omega$ and polarisation $j$ (s - TE polarisation; p - TM polarisation).
The transmission factor   $\mathcal{T}_j (\omega, \mathbf{k}_\perp;d)$ is fully determined by the optical properties of the receiver
and emitter. The contributions of the propagating waves are restricted to waves with wavevectors $|\mathbf{k}_\perp| \leq k_0$ and the 
evanescent wave contributions stem from waves with wavevectors $|\mathbf{k}_\perp| > k_0$.

\begin{figure}
  \includegraphics[ width = 0.45\textwidth]{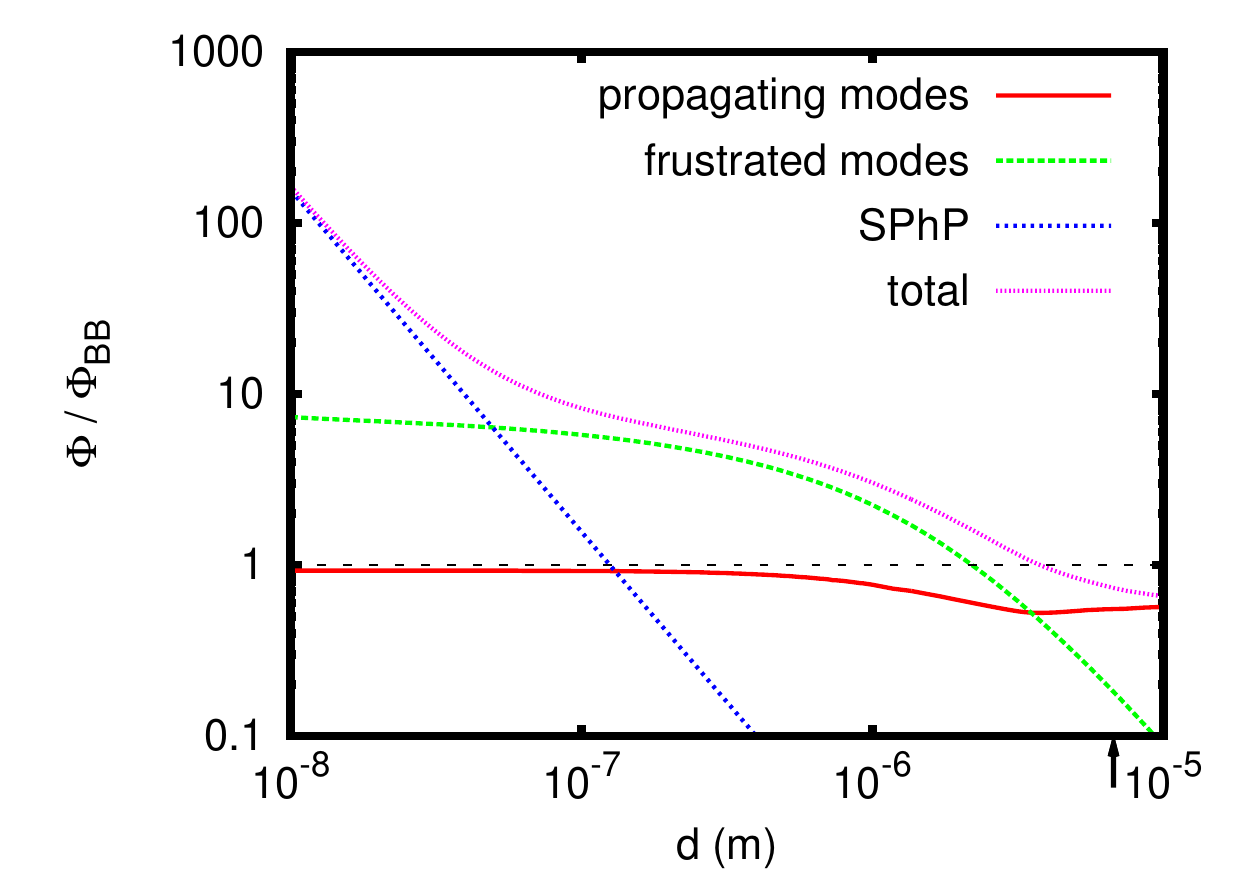}
  \caption{Near-field heat flux $\Phi$ between two hBN halfspaces normalized to the blackbody value $\Phi_{\rm BB}$ as function of distance $d$. Here the temperature of the emitter is 300K and the temperature of the receiver is 0K so that $\Phi_{\rm BB} = 459.27\,{\rm W}{\rm m}^{-2} = 0.046\,{\rm W}{\rm cm}^{-2}$. The contribution of different modes to the near-field heat flux is highlighted. For more details see Ref.~\cite{Intech}.~\label{Fig:DiffModes}}
\end{figure}

It is clear that the maximum contribution of the propagating waves is obtained by setting $\mathcal{T}_j \equiv 1$ for all  $|\mathbf{k}_\perp| \leq k_0$. 
Then the above expression gives the blackbody result
\begin{equation}
   \Phi_{\rm BB} = \sigma_{BB} \bigl[ T^4_{\rm e} - T^4_{\rm r} \bigr]
\end{equation}
where $\sigma_{BB} = 5.67\times 10^{-8} {\rm W} {\rm m}^{-2} {\rm T}^{-4}$ is the Stefan-Boltzmann constant. As a consequence it becomes 
apparent that the contribution of the evanescent waves will add to the heat flux of the propagating waves and therefore it can 
be larger than $\Phi_{\rm BB}$. In Fig.~\ref{Fig:DiffModes} the contributions of the different heat flux channels for two hBN plates is highlighted. 
It can be seen that the propagating modes dominate in the far-field regime with $d > \lambda_{\rm th}$ whereas the evanescent modes dominate in the 
near-field regime with $d < \lambda_{\rm th}$. The near-field regime can be devided into the regime $d < \lambda_{\rm th}/10$ where 
the surface modes dominate and the intermediate regime with $\lambda_{\rm th}/10 < d <  \lambda_{\rm th}$ where the frustrated modes 
dominate~\cite{Volokitin,Pendry,Intech}. The behaviour observed Fig.~\ref{Fig:DiffModes} is similar for other materials like 
SiC, silica or saphire supporting surface modes in the infrared. For example, for SiC with a surface mode resonance at $\omega_{sp} = 1.787\times10^{14}\,{\rm rad/s}$ the enhancement would be 1000$\Phi_{\rm BB}$ at $d = 10{\rm nm}$ for the choice of temperatures in Fig.~\ref{Fig:DiffModes} and for silica even higher enhancement factors are achievable. Depending on the shape and temperature regimes optimized material parameters can be obtained to increase the energy density and the heat flux~\cite{Miller, Miller2,Alejandro}. 

First measurements of the near-field heat transfer between two planar samples were already conducted in the 1970's by 
 Hargreaves and Domoto {\itshape et al.}~\cite{CravalhoExp,Hargreaves,Domoto2} for metallic samples. 
A new era of near-field measurements has been started with the seminal work of Ottens {\itshape et al.} in 2011~\cite{Ottens} who could measure the heat flux between 
macroscopic $50{\rm mm}$\,x\,$50{\rm mm}$ saphire plates, which support surface modes in the infrared, down to a distance of about 2 micrometers. 
The aspect ratio between the smallest interplate distance to the lateral extension of the samples is approximately 1:25000. The measured enhancement was found to be 1.26 times the 
blackbody value~\cite{Ottens} and can be attributed to the contribution of the frustrated total reflection modes.  The following works have 
improved the experimental techniques in order to reduce the interplate distances as much as possible and to achieve 
the highest possible near-field enhancement~\cite{Ottens,Hu,Kralik,Gelais,Song,Bong,Ito,Watjen,Song2,Bernadi,Lang,Ghashami,Lim,Reddy}. State-of-the art setups like the experiment by Reddy's group in Ref.~\cite{Reddy} are now capable of measuring the heat flux between a $50\mu{\rm m}$\,x\,$50\mu{\rm m}$ silica emitter and an extended silica receiver down to 25nm and found an enhancement of about 700 times the blackbody 
value proving the possibility to have extremely large radiative heat fluxes due to the surface mode interaction in silica. Note that in 
most recent experimental setups the smaller distances are achieved by reducing the size of the samples so that for example the aspect ratio in 
the experiment~\cite{Reddy} is 1:2000. 

One of the technical challenges for near-field thermophovoltaics will be to use the near-field effect for large scale PV cells. 
To achieve this coal several structures using different kind of spacers having a small heat conductivity were proposed~\cite{Hu,Ito,Bernadi,Lang}. 
The spacers make it possible to keep the emitter and receiver in a controlled near-field distance. For example, Ito {\itshape et al.} measured in such a structure the radiative heat flux between two $19{\rm mm}$\,x\,$8.6{\rm mm}$ quartz plates with microstructured pillars down to 500nm~\cite{Ito}, i.e.\ the aspect ratio is in this case about 1:17200. One of the drawbacks of such a solution is that the heat conduction through the pillars is relatively large. Consequently by this conduction the receiver will heat up which will reduce the efficiency of any near-field thermophotovoltaic device. In the setup of Ito {\itshape et al.} the heat flux through the spacers was on the same order of magnitude as the radiative heat flux between the emitter and the receiver. The goal of future setups or devices using spacers is therefore clearly to reduce this contribution of heat conduction through the spacers while having large areas for thermal radiation and a high stability.

\section{Near-field energy conversion}

As demonstrated in the previous sections the radiative heat flux by the surface-phonon-polarition  contribution can be extremely large 
and as the surface-phonon-polariton contribution to the energy density it can also be quasi-monochromatic. This is a promising feature 
for an exploitation in near-field 
thermophotovoltaic devices. It has been shown in a theoretical study by Narayanaswamy and Chen in 2001~\cite{Narayanaswamy} that the 
surface modes of phonon-polaritonic emitters like SiC, cubic and hexagonal boron nitride will result in a tremendous increase of the 
heat flux into a direct band gap PV cell. A first quantitative calculation of the photocurrent and electric power in a tungsten-GaSb 
configuration was provided by Laroche {\itshape et al.}~\cite{Laroche} assuming 100\% quantum efficiency of the PV cell, i.e.\ it is assumed
that all the incoming or absorbed heat radiation will be converted into a photocurrent. It could be shown that the conversion efficiency
converges in this case towards 29\% for distances below 100nm. Theoretically a generated electric power output of $3\times10^4\,{\rm W} {\rm m}^{-2}$ was 
reported for a distance of 5nm. It should be noted that the Shockley-Queisser limit for a blackbody illuminating a GaSb 
cell is an efficiency of 29\%. It was theoretically demonstrated that a quasi-monochromatic emitter like those considered by 
Narayanaswamy and Chen~\cite{Narayanaswamy} can even beat the Shockley-Queisser limit in the near-field regime yielding 35\% at an extremely
small distance of 5nm which can be hardly achieved in real NTPV devices.

\begin{figure}
 \includegraphics[ width = 0.55\textwidth]{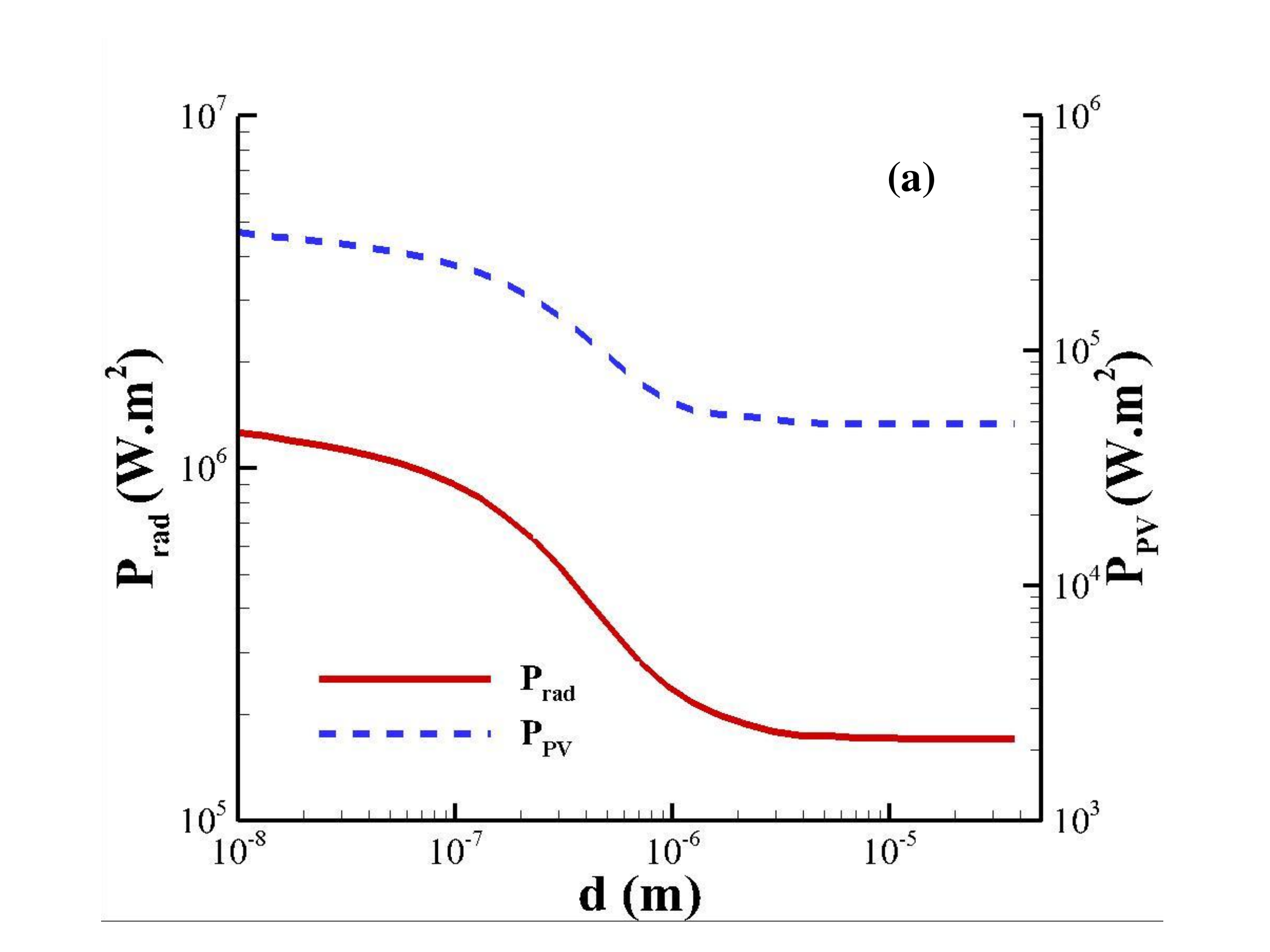}
 \includegraphics[ width = 0.55\textwidth]{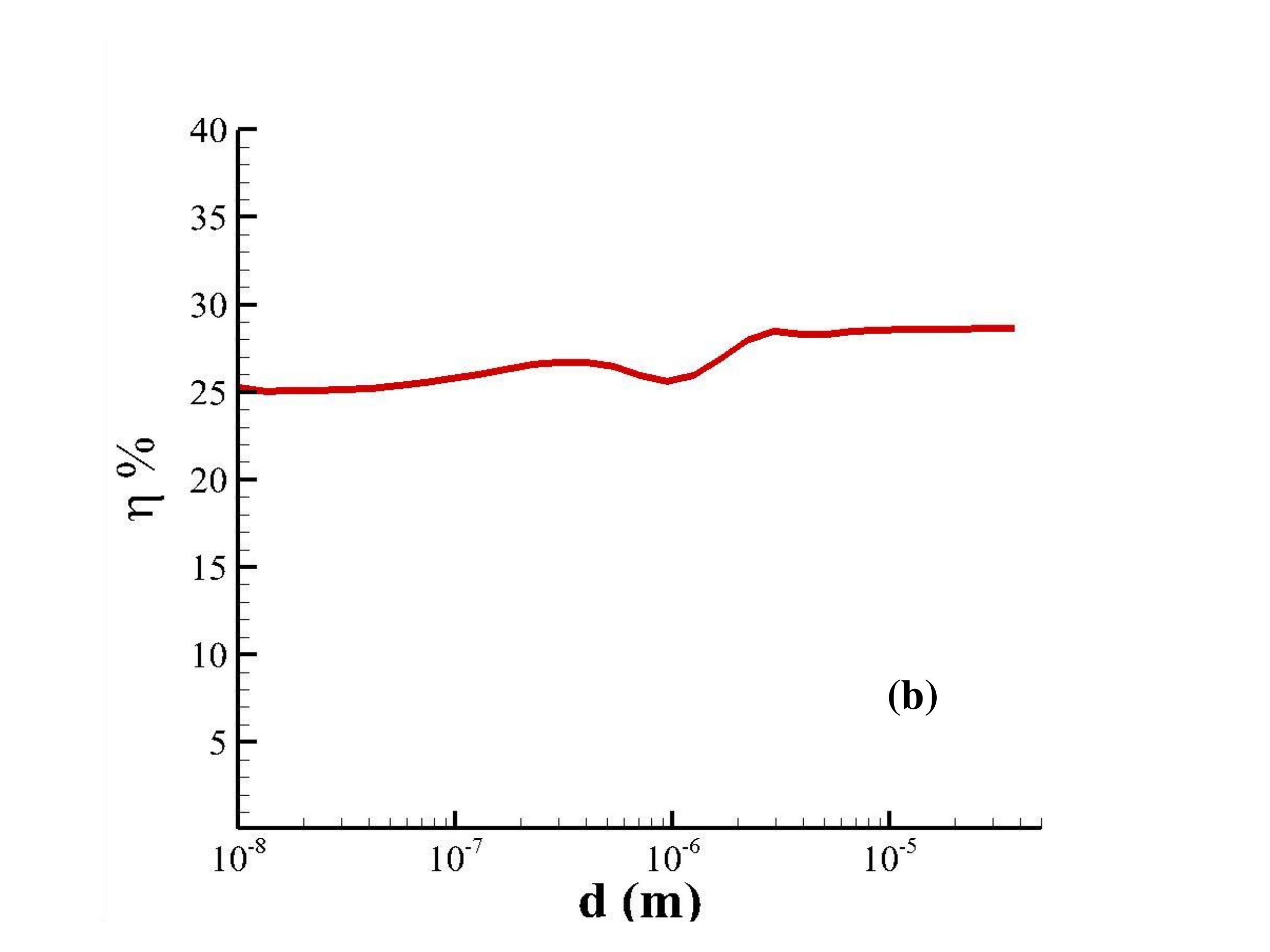}
  \caption{(a) Radiative power exchanged between a $hBN$ thermal source at $T_{\rm e} = 1500\:{\rm K}$ and a InSb junction at $T_{\rm s}=300\:{\rm K}$ and electric power generated in this system with respect to the separation distance. (b) Efficiency of produced electricity in a hBN/InSb TPV conversion system with respect to the separation distance between the source and the cell.}
\end{figure}

To illustrate the potential of the NTPV technology for energy harvesting we consider a system composed by a hot emitter made of hexagonal Boron Nitride (hBN) at temperature $T_{\rm e}$ placed in the proximity of a junction cell at temperature $T_{\rm r}<T_{\rm e}$ made of Indium Antimonide (InSb) with a gap frequency $\omega_{g}=1.8231\times 10^{14} {\rm rad/s}$ ($E_g = 0.12\,{\rm eV}$). We use the optical properties of hBN from Eq.~(\ref{Fig:hBN}) and the dielectric function of the PN junction defined as~\cite{Forouhi}
\begin{equation}
\begin{split} 
  \epsilon_{\rm r} (\omega)= \begin{cases} n_r^2(\omega), & \omega < \omega_g,\\
   (n_r(\omega)+ {\rm i} n_i(\omega))^2, & \omega > \omega_g, \end{cases}\\
\label{epsilon_cell}
\end{split} 
\end{equation}
where 
\begin{equation}
  n_r=\underset{j=1}{\overset{4}{\sum}}\frac{B_j\omega+C_j}{\omega^2-B_j\omega+C_j}
\end{equation}
and 
\begin{equation}
   n_i=\frac{\alpha_0 c}{2\omega}\sqrt{\frac{\omega}{\omega_g}-1}
\end{equation}
with $B_j=\frac{a_j}{q_j}(E_g b_j-\frac{1}{2}b_j^2-E_g^2+c_j)$, $C_j=\frac{a_j}{q_j}((E_g^2+c_j)\frac{b_j}{2}-2E_g c_j)$ and $q_j=\frac{1}{2}\sqrt{4c_j-b_j^2}$. The constants $a_j$, $b_j$ and $c_j$ ($j = 1-4$) are tabulated values and dependend on the type of semiconductors~\cite{Forouhi}. It is direct to verify that hBN supports a surface phonon-polariton resonance at frequency $\omega_\text{spp}\simeq2.960\times10^{14}\,\text{rad s}^{-1}$, larger than $\omega_g$ as desired.

The radiative power exchanged between the source and the junction reads
\begin{equation}
\begin{split}
  P_{\rm rad} &= \int_0^\infty\!\frac{d\omega}{2\pi} K(\omega) [\Theta(\omega,T_{\rm e})\\
   &\qquad\qquad- \Theta(\omega-\omega_{0},T_{\rm r}) H(\omega-\omega_{g})].
\end{split}
\label{Eq:Pradi}
\end{equation}
where $H$ denotes the Heaviside function, $\omega_{0}=e V_{0}/\hbar$. $V_{0}$ being the potential difference at which the  cell is operating and which is here~\cite{Messina} taken as $\hbar \omega_g (1 - T_{\rm r}/T_{\rm e}) / e$ introducing the elementary charge $e$. The quantity
\begin{equation}
  K(\omega)= \sum_{j = \{\rm s,p\}} \int\!\frac{\mathbf{d^2 \mathbf{k}_\perp}}{(2\pi)^2} \mathcal{T}_j(\omega,\mathbf{k}_\perp;d)
\end{equation}
is the number of modes per unit area which contribute to the transfer weighted by the Landauer transmission probability  $\mathcal{T}_j(\omega,k)$  between  the source and the cell in polarization $j$ of the energy $\hbar \omega$ carried by the mode $(\omega,\mathbf{k}_\perp)$. The two terms in expression (\ref{Eq:Pradi}) represent the emission radiated by the source toward the cell and the power radiated back from the cell in direction of the source, respectively. 
As for the electric power density which is generated in the cell, it reads
\begin{equation}
   P_{\rm PV}=e V_{0}\int_{\omega_{g}}^\infty\!\frac{d\omega}{2\pi} [\Theta(\omega,T_{\rm e})-\Theta(\omega-\omega_{0},T_{\rm r})] \frac{K(\omega)}{\rm \hbar \omega}
\label{Eq:Ppvi}
\end{equation}
assuming a quantum efficiency of 100\%.
In Figure 5  we show the evolution of the power transferred to the cell and the electric power generated by the system with respect to the separation distance $d$ between the source and the  cell. We also plot in Fig. 5(b) the efficiency $\eta = P_{PV}/P_{rad}$ of the cell. For separation distances larger than the thermal wavelength $\lambda_{th}$ the power exchanged between the source and the cell is mainly due to propagating photons so that it is independent on the distance. At subwavelength distances the radiative heat exchanges increase as the separation distance decays. This enhancement of exchanges is directly related to the number of contributing modes with respect to the separation distance~\cite{LandauerPBA,LandauerSAB}. As discussed in the previous section at subwavelenth distance the non-propagating modes (i.e. evanescent modes)  superimpose to the propagating ones giving rise to new channels for heat exchanges by photon tunneling between the source and the cell. As a direct consequence these photons can generate further electron-hole pairs inside the junction and therefore increase the production of electricity. Hence at $d=500\,{\rm nm}$ the generated power is doubled compared to what happens in far-field regime. In extreme near-field regime this power is almost an order of magnitude larger than in far-field. As the efficiency is concerned we see in Fig.5(b) that in near-field the efficiency is a bit reduced  in comparison with the  performances of TPV conversion device in  far-field regime. Nevertheless we have in this case an efficiency of about 25\% and the output power at a realistic distance of 100nm is approximately $220\,{\rm kW}{\rm m}^{-2}$. Note that while the surface phonon-polariton of hBN is located at 5$\mu m$ the gap of InSb cell is around 7.3 $\mu m$. This spectral mismatch reduces the conversion performances . By adding a sheet of graphene to the emitter or receiver, the heat flux by the surface modes and hence the efficiency can be increased in the extreme near-field regime~\cite{Messina,Ilic,23}. In Ref.~\cite{Ilic} an efficiency of 30\% (40\%) with a generated power of 6 ${\rm W}{\rm cm}^{-2}$ (120 ${\rm W}{\rm cm}^{-2}$)  was reported for an emitter temperature of 600K (1200K) at a distance of 10nm.

\section{State of the art}

By using a much more realistic model of the PV cell Park {\itshape et al.}~\cite{Park} have shown that in a tungsten-InGaSb configuration
the efficiency can substantially decrease when making the distance between the emitter and the PV cell smaller. This is due to the
fact that the penetration depth of the evanescent waves inside the PV cell becomes very small, becaue the evanescent waves are exponentially
damped on a distance on the order of the distance $d$ in the quasi-static regime. For surface modes it could be shown~\cite{Basu} 
that the penetration depth scales like $0.25 d$ for phonon-polaritonic materials like SiC. As a consequence, for very small distances the heat 
flux increases dramatically due to the evanescent wave contributions, but it will be dissipated at the interface mainly. Hence, only electron-hole 
pairs close to the interface of the PV cell will be generated limiting such the conversion efficiency. Park {\itshape et al.} report an efficiency 
of about 20\% at a distance of 10nm. Nonetheless, the theoretically predicted generated electric power is still high and has a nominal value 
of $10^6\,{\rm W}{\rm m}^2$. Introducing a back reflector to the PV cell could even improve the performance~\cite{BasuReview}. A detailed discussion of
the contribution of the different heat flux channels to the NTPV efficiency can be found in Ref.~\cite{Bernadi2}. 

Since the work of Park {\itshape et al.} the modelling has been further improved and the Moss-Burnstein effect, the impact of series resistance, photon recycling as well as parasitic sub-bandgap absorption and cell cooling has been studied in detail~\cite{Fan, Blandre, DeSutter, Vaillon,Mirmoosa1}. Furthermore, hyperbolic materials have been proposed for replacing the emitter, receiver or the gap region~\cite{Nefedov, Simovski, Jin, Mirmoosa2}, because with these materials spectral control, high heat flux levels, and large penetration depths are achieavable~\cite{Jin,Biehs, Biehs2,Shen,Lang2,Tschikin}. For example, with the NTPV system in Ref.~\cite{Jin} using a W/SiO$_2$ hyperbolic emitter and a InAs cell a generated power of $1.78{\rm kW}{\rm m}^{-2}$ is reported at a realistic distance of 100nm.

Despite all those efforts on the theoretical side there is only a single up-to-date experimental study of a NTPV system conducted by 
Reddy's group~\cite{ReddyNTPV}. In this pioner experimental study a Si emitter at temperatures ranging from $525\,{\rm K}$ to $665\,{\rm K}$ is brought in close 
vicinity of two different TPV cells with band gaps of $0.345\,{\rm eV}$ and $0.303\,{\rm eV}$. The measured power generated at the high (small) band-gap cell 
at a nominal distance of 60nm is 40 times larger than at long separation distances. However this power remains relatively small (around 30nW  for $T_{\rm e} = 655\,{\rm K}$) because heat is essentially  transfered with frustrated photons.
Taking into account the area of circular emitter of radius of $40\,\mu {\rm m}$ we obtain a power flux of about $6\,{\rm W} {\rm m}^{-2}$ and an aspect ratio of 1:3200. Although this value is relatively small and corresponds to an extremely low conversion efficiency ($\approx 0.02\%$) several points can significantly be  improved. In particular by scaling up the system size and increasing the emiter temperature $1000\,{\rm K}$ a 6\% conversion efficiency can be expected. Also using an emitter which support a surface wave in the Planck window the radiative heat exchanges between the emitter and the PV cell could surpass by several orders of magnitude the flux exchanged at long separation distances. With these improvements the efficiency of NTPV systems could  be even larger than the Schockley-Queisser limit~\cite{Shockley}.

\section{Concluding remarks}

In conclusion, the basic concepts to convert the near-field electromagnetic energy confined close to a hot body into electricity  have been introduced. We have shown that  this near-field technology allows to largely surpass the performances of the classical TPV technology. Despite of its potential several hurdles strongly limit to date the development and the massive deployment of this technology. On the one hand, near-field heat flux experiments between planar structures in the near-field regime are nowadays possible even down to 25nm distance, but typically not for systems with large areas. Using low conducting spacers to keep the vacuum gap in the system is one possible work around. Another is to replace the vacuum gap by a low conducting but high index or hyperbolic material as recently proposed~\cite{Mirmoosa2}. Furthermore, it is not trivial to maintain large temperature gradients over very small gaps and the cooling of the cell could further reduce the efficiency of the cell, because the power used to cool the cell has to be taken into account in a global balance. So far, a power output of only $6\,{\rm W} {\rm m}^{-2}$ and an efficiency of only 0.02\% seems to be a little bit disappointing, but the NTPV technology has a great potential and we are very optimistic that future theoretical and experimental works will improve the efficiencies of NTPV systems step by step to a point where the promissed near-fied enhancement will be achieved and power outputs of $1{\rm kW}\,{\rm m}^{-2} - 1 {\rm MW}\,{\rm m}^{-2}$ and efficiencies around 20-40\% are not just the dream of theorists but reality.

\acknowledgments
\noindent

S.-A.\ B. acknowledges support from Heisenberg Programme of the Deutsche Forschungsgemeinschaft (DFG, German Research Foundation) under the project No. 404073166.

\end{document}